\documentclass[11pt]{article}
\usepackage{newpasp_astroph,graphics,natbib}

\index{Mamon, G. A.}

\begin{document}

\title{Theory of galaxy dynamics in clusters and groups}

\author{Gary A. Mamon}
\affil{Institut d'Astrophysique, F--75014 Paris, FRANCE} 





\begin{abstract}
Analytical estimates of the mass and radial dependence of the
rates of galaxy mergers and of tidal
interactions are derived for clusters and groups of galaxies, taking into
account the
tides from the system potential that limit the sizes of galaxies.
Only high mass galaxies undergo significant major merging before being
themselves 
cannibalized by more massive galaxies.
Strong tides from the group/cluster potential severely limit the merger/tide
cross-sections in the central regions, and while tides are most efficient at
the periphery, one should see merging encounters further inside rich clusters.

\end{abstract}




\section{Introduction}

Mergers of galaxies in slow collisions and tidal interactions in rapid
collisions are two key dynamical processes that occur in groups and clusters
of galaxies.
Cosmological $N$-body simulations are beginning to approach the
resolution necessary to study galaxy dynamics in groups and clusters (see
Moore, in these proceedings).
Moreover, mergers and tidal collisions leave significant observational
signatures, in the form of tidal tails, asymmetries and generally disturbed
morphologies and internal kinematics (see Amram, in these proceedings).
Also, galaxy merging is an essential mechanism for driving elliptical
galaxy morphologies given disk-like progenitors.
As such, an understanding of galaxy merging is very important for
semi-analytical modeling of galaxy formation.

In this review, I compute analytically
the rates at which a galaxy of given mass and
position within a cluster or group with a \citeauthor*{NFW95}
(\citeyear{NFW95}, NFW) potential 
undergoes slow major mergers with lower
mass galaxies and rapid tidal encounters.
Since the collision cross-sections are strongly modulated by the tides from
the group/cluster potential, I begin with a simple formalism for estimating
the tides from the system potential.
Note that I will not consider ram pressure stripping on galaxies.

\section{Tides from the cluster/group potential}

The potential of a cluster can exert a strong differential force on a galaxy
orbiting within it, but these tides are strongly dependent on the galaxy
orbit.

A galaxy on a nearly
circular orbit is likely to be tidally locked, as the Moon is
with respect to the Earth.
In this case, the tidal force is simply \citep{King62}
\begin{equation}
F_{\rm tide} = \Delta \left [ {GM(R) \over R^2} - \Omega^2(R) \right ]
\label{ftide}
\end{equation}
and equating $F_{\rm tide}$ in eq.~(\ref{ftide})
to the force, $f = Gm(r)/r^2$ that a galaxy
exerts on one of its 
stars yields for $r \ll R$ a galaxy tidal radius $r_t$ that satisfies a
velocity modulated density criterion (see \citealp{M95_Chalonge}):
\begin{equation}
\bar \rho_g (r_t) = \bar \rho_{\rm cl} (R) \left [ 2 - {3 \rho_{\rm cl} (R)
\over \bar \rho_{\rm cl} (R) } + {V_p^2 (R) \over V_{\rm circ}^2(R)} \right ]
\ ,
\label{circtide}
\end{equation}
where $V_p$ and $V_{\rm circ}$ are respectively the galaxy's velocity at
pericenter and the cluster's circular velocity.
The term in brackets in eq.~(\ref{circtide}) is 2 for circular orbits.
Moreover, for a singular isothermal law $\rho \sim r^{-2}$ for both galaxy
and cluster, one finds 
(see \citealp{White83}) that for circular orbits
the galaxy size is proportional to its clustocentric
radius: 
$r_t = 2^{-1/2} (v_{\rm circ} / V_{\rm circ}) R$, where $v_{\rm circ}$ is the
circular velocity of the galaxy.
For general density profiles, writing
$\bar \rho (r) 
\sim v_{\rm circ}^2(r)/r^2$, one obtains
\begin{equation}
{r_t/R \over v_{\rm circ} (r_t) / V_{\rm circ} (R)} = \left (2 -
3\,{\rho \over \bar \rho} + V_p^2/V_{\rm circ}^2(R) \right)^{-1/2} \ .
\label{rtcirctide}
\end{equation}
\cite{Merritt84} used a similar circular-tide criterion to argue that
galaxies are strongly tidally limited by the cluster potential.

However, cosmological infall imposes elongated orbits.
A galaxy on an elongated orbit experiences a strong tide during its rapid,
hence short, passage at pericenter, prompting \cite*{OSC72} to introduce the
term {\sl tidal shock\/}.
During this shock, a star in the galaxy experiences a velocity impulse
\begin{equation}
\Delta v \sim F_{\rm tide} \Delta t \sim {GM(R_p) \,r \over R_p^3} \left ({R_p
\over V_p} \right ) = {\rm cst} {GM(R_p) \,r \over R_p^2 V_p} \ ,
\label{dvimpulse}
\end{equation}
where we neglected the centrifugal term in $F_{\rm tide}$, because the galaxy
falls in too fast to be phase locked.
A more precise calculation by \cite{Spitzer58}, who introduced 
the {\sl impulsive approximation\/}
where the point-mass perturber
moves at constant ${\bf V}$, produces the same relation as in
eq.~(\ref{dvimpulse}) with a constant of order unity.
The impulse approximation can also be applied to extended perturbers
\citep{AW85,M87}. 
Recently, \cite*{GHO99} applied the impulsive approximation to the more
realistic \cite{Hernquist90} 
potential and found a dependence of $\Delta v$ matching that of
eq.~(\ref{dvimpulse}), with a very small dependence on $R_p$ and again the
constant is found to be of order unity for elongated orbits (with order of
unity changes when they performed orbit-integrated --- instead of
straight-line --- tidal calculations).

The tidal radius can then be defined as that where the energy increment
caused by the tidal perturbation is equal to the binding energy
\citep{White83}. 
With $E \sim Gm(r_t)/r_t$ and $\Delta E \sim (\Delta v)^2/2$, one obtains
another velocity modulated density
criterion
\begin{equation}
\bar \rho_g(r_t) \simeq \bar \rho_{\rm cl} (R_p) \left ({V_{\rm circ}(R_p)
\over V_p} \right )^2 \ ,
\label{elongtide}
\end{equation}
which for any density profile yields
\begin{equation}
{r_t/R \over v_{\rm circ} (r_t) / V_{\rm circ} (R)} = {V_p \over V_{\rm
circ}} \ .
\label{rtelongtide}
\end{equation}

\begin{figure}[ht]
\begin{center}
\resizebox{0.7\hsize}{!}{\rotatebox{-90}{\includegraphics{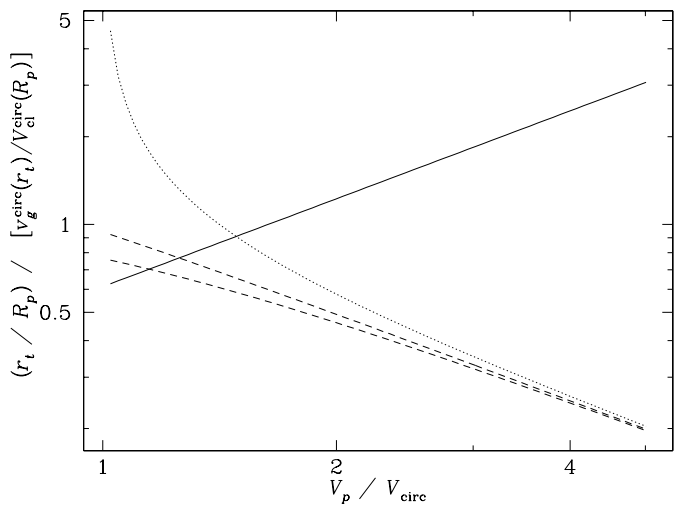}}}
\end{center}
\caption{
Velocity modulation of normalized tidal radii for given orbit pericenter.
The \emph{solid curve} shows impulsive tides for elongated orbits
(eq.~[\ref{rtelongtide}]).
The \emph{dashed curves} show circular tides (eq.~[\ref{rtcirctide}])
in a
\citeauthor*{NFW95} potential for 0.1 (\emph{upper curve}) and 1 (\emph{lower
curve}) scale radii.
The \emph{dotted curve} shows the circular tide within a homogeneous core.
}
\label{tides}
\end{figure}
Figure~\ref{tides} shows the effects of velocity modulation on the tidal
radii. 
The effective tidal radius should be taken as the largest of the circular and
impulsive regimes (otherwise one would be left with discontinuities in the
transition $V_p$ from near-circular to elongated orbits).
Hence, with homogeneous cores, circular tidal theory produces increasingly
smaller tidal radii for orbits of increasing but low elongation, as
\cite{MW87} found in their $N$-body simulations.
With cuspy cores as in the NFW
profile, low-elongation non-circular
orbits experience tidal shocks instead. 
The orbit elongation, $R_p/R_a \simeq 0.2$, found in \citeauthor{Ghigna+98}'s
(\citeyear{Ghigna+98}) 
high-resolution cosmological simulations corresponds to $V_p/ V_{\rm
circ}(R_p) \simeq 1.5-2.7$, roughly yielding
\begin{equation}
r_t \simeq (1.2-1.5) \,
R_p \left ({v_{\rm circ} \over V_{\rm circ}} \right) \approx (1.2-1.5)\,R_p 
\left ({v_g \over v_{\rm cl}}\right)
\label{finalrt}
\end{equation}
as for singular isothermal
models, where $v_g$ and $v_{\rm cl}$ are the mean
galaxy and cluster velocity dispersions, respectively.
If $V_p \simeq V_{\rm circ} (R_p)$ and 
if galaxy and cluster density profiles are self-similar,
then tides from the cluster potential would
force the simple relation $m(r_t)/m(r_{\rm vir}) = M(R_p)/M(R_{\rm vir})$.
In fact, using $V_p/V_{\rm circ}(R_p)$ expected for the NFW profile, assuming
$\langle R \rangle \simeq 4 R_p$, and adopting the departures from
self-similarity in the NFW profiles noted by \citeauthor*{NFW97}
(\citeyear{NFW97}, lower mass NFW
profiles are more centrally concentrated), one finds 
\begin{equation}
{m(r_t) \over m(r_{\rm vir})}
\simeq \left [{M(R) \over M(R_{\rm vir})}\right]^{b_m} \simeq a_r \,
\left ({R \over R_{\rm vir}} \right)^{b_r} \ ,
\label{mtide}
\end{equation}
where for clusters ($v_{\rm cl} = 1000 \, \rm km \, s^{-1}$) and
groups ($v_{\rm cl} = 300 \, \rm km \, s^{-1}$) we respectively have
$b_m = 0.50$ and 0.82,
$a_r = 0.58$ and 0.52, and $b_r = 0.57$ and 0.78
(eq.~[\ref{mtide}] is accurate to better than 5\% for $R > 0.05\,R_{\rm vir}$).

\section{Galaxy merger rates in clusters and groups}

\subsection{Global merger rates for equal mass galaxies}

The rate of mergers
is obtained by integrating over velocities the merger cross-sections:
\begin{equation}
k \equiv {1\over n^2} \,{d^2N \over dt\,dV} = 
\langle v s(v) \rangle = \int_0^\infty dv f(v) v s(v) \ ,
\label{genmergerrate}
\end{equation}
where $s(v) = \pi [p_{\rm crit} (v)]^2$ is the merger cross-section and
$f(v)$ is 
the distribution of 
relative velocities (with $\int_0^\infty f(v) \,dv = 1$).
Hence, $nk \equiv dN/dt$ is the rate at which a galaxy suffers a merger.
Within the virialized regions of clusters with 1D velocity dispersion $v_{\rm
cl}$, the velocity distribution is  
a gaussian with standard deviation $2^{-1/2} v_{\rm cl}$:
$f(v) = 2^{-1} \pi^{-1/2} v_{\rm cl}^{-3} v^2 \exp[-v^2/(4\,v_{\rm cl}^2)]$.

\cite{RN79}, \cite{AF80} and \cite{FS82} 
have established merger cross-sections from very
small $N$-body
simulations of galaxy collisions, that were based upon the parameters at
closest approach.
The maximum distance of closest approach, $r_p^{\rm max}$, was 4
(\citeauthor{AF80}) or 11 (\citeauthor{FS82}) times the mean
galaxy half-mass radius, $r_h$.
\cite{M92} used the \citeauthor{RN79} cross-section with the
\citeauthor{AF80} scaling to derive a merger rate.

However, the gaussian approximation for the relative velocity distribution
implies that the
cross-sections used in eq. (\ref{genmergerrate}) are based upon impact
parameters (at infinity) and not at closest approach.
\cite{MH97} derived merger cross-sections using high resolution
$N$-body 
simulations of colliding galaxies with more realistic density profiles.
Their cross-sections are expressed in terms of impact parameters.

\cite{KK97} used a simple gravitational focusing recipe to connect the
cross-section at closest approach (which they assumed to be independent of
pericentric velocity), in units of the velocity at infinity,
to the cross-section at infinity (without making any assumption on the
potential energy  of interaction of the colliding pair).
Note that \citeauthor{MH97} show that $r_p^{\rm max} > 10\,r_h$, while
\citeauthor{KK97} 
argue that it is approximately the sum of the galaxy \emph{radii} (which are
ill-defined).

\begin{figure}[ht]
\begin{center}
\resizebox{0.7\hsize}{!}{\rotatebox{-90}{\includegraphics{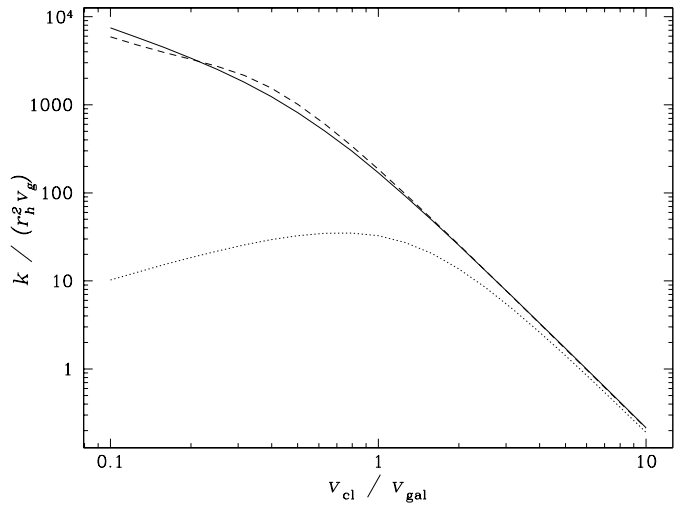}}}
\end{center}
\caption{Dimensionless merger rates $k/(r_h^2 v_g)$
(see eq.~[\ref{genmergerrate}])
as a function of the ratio of 
cluster to galaxy velocity dispersion.
The \emph{dotted}, \emph{solid} and \emph{dashed} curves respectively 
correspond to the
rates of \cite{M92}, \citeauthor{MH97} (\citeyear{MH97}, using
\citealp{Hernquist90} model galaxies) and 
\citeauthor{KK97} (\citeyear{KK97}, rescaled vertically, because they reason
in terms of galaxy 
radii rather than galaxy half-mass radii).
}
\label{globmergrate}
\end{figure}

Figure~\ref{globmergrate} shows the dimensionless merger rates
$k /( r_h^2 v_g)$, where $v_g$ is the mean galaxy internal velocity dispersion,
 derived from
\cite{M92}, \cite{MH97}, and \cite{KK97}. I rescale the rates of
\citeauthor{KK97} in terms of half-mass radii using $R = 9\,r_h$ (to obtain
 merger rates similar to those of \citeauthor{MH97}), {\it i.e.\/,} $r_p^{\rm
max}/r_h = 18$.

The agreement between the merger rates of \citeauthor{KK97} and
\citeauthor{MH97} is remarkable, given the very simple analytical formulation
of the former authors (but again this required a rescaling, or, in other
words, a choice of $r_p^{\rm max}/r_h = 18$).
The very low merger rate of \cite{M92} in the group regime ($v_{\rm cl}
\approx v_g$) is a consequence of the lack of gravitational focusing in
that model.

The merger rates of \cite{M92} and \citeauthor{MH97} agree to within 15\% in
the cluster regime ($v_{\rm cl} \ga 4 \,v_g$). This agreement is almost
fortuitous since \citeauthor{M92} shows that the merger rate with the
\citeauthor{RN79} cross-section scales as $(r_p^{\rm max} / r_h)^2$, while
this ratio is very different in \citeauthor{MH97}'s cross-section.
In any event, in the cluster regime the merger rate can then be written
\begin{equation}
k = b \,{r_h^2 v_g^4 \over v_{\rm cl}^3} = a \,{G^2 m^2 \over v_{\rm cl}^3} \ ,
\label{clustermergrate}
\end{equation}
where $a \simeq 8$ \citep{M92}. 
With $3 \,v_g^2 \simeq 0.4 \,G m/r_h$
\citep{Spitzer69}, appropriate for the \citeauthor{Hernquist90} model, the
\citeauthor{MH97} rate translates to $a = 12$.
Figure~\ref{globmergrate} shows that $k \sim v_{\rm cl}^{-3}$, whichever merger
cross-section is used.
In fact, it is easy to show that \emph{for any merger cross-section rapidly 
decreasing with increasing velocity}, the merger rate should scale as $v_{\rm
cl}^{-3}$ for $v_{\rm cl} \gg v_g$, as first found by \cite{M92} for the
\citeauthor{RN79} cross-section.

The important conclusion of Figure~\ref{globmergrate} is that \emph{for given
galaxy parameters, the merger rate is roughly 100 times lower in rich clusters
than in poor groups of galaxies}.

\subsection{Merger rates for different masses}

If the critical merging velocity $v_{\rm crit}$
is a function of $r_p/r_h$ \citep{AF80,FS82}, it is easy to
show that $k \sim r_h^2$, and if $v_{\rm crit}$ is a function of $r_p/\langle
r_h \rangle$, then $k \sim \langle r_h \rangle^2$ (see \citealp{M92}).
Similarly, it is reasonable to expect that $k \sim \langle v_g^2 \rangle^2$.
Then, given eq.~(\ref{clustermergrate}) and that $m \sim r_h^3$, 
the rate of mergers of a galaxy of mass $m$ with
a galaxy of mass $\lambda m$ 
\begin{equation}
k(m,\lambda m) = {a G^2 m^2 \over v_{\rm cl}^3} 
\left ( {1 + \lambda^{1/3} \over 2} \right )^2  
\,
\left ( {1 + \lambda^{2/3} \over 2} \right )^2  
\ .
\label{nkmmgen}
\end{equation}

A given galaxy undergoes mergers with other galaxies at a rate
\begin{equation}
{\cal R} \equiv
n \bar k(m) = \int_{\lambda_{\rm min}}^{\lambda_{\rm max}} k(m,\lambda m)
\,n(\lambda m)\, d(\lambda m) \ ,
\label{nkgen}
\end{equation}
where for major mergers with smaller galaxies (that transform disk galaxies
into ellipticals), $\lambda_{\rm min} \simeq 1/3$ and $\lambda_{\rm max} =
1$, while for destruction by mergers with larger galaxies, $\lambda_{\rm min}
= 1$ and $\lambda_{\rm max} \to \infty$.
Adopting a \cite{Schechter76} form for the mass function of galaxies, $n(m) =
(n_*/m_*)\, x^{-\alpha} \exp(-x)$, where $x = m/m_*$, 
eqs.~(\ref{nkmmgen}) and
(\ref{nkgen}) yield
\begin{equation}
{\cal R} = n \bar k = {a G^2 n_* m_*^2 \over 16 \,v_{\rm cl}^3} \,K(m/m_*) \ ,
\label{nkofm} 
\end{equation}
\begin{equation}
K_{\rm major} (x) = 
x^{3-\alpha} \sum_{j=0}^6
{\rm Min} (j,7\!-\!j) \,\left [
\Gamma(1\!+\!j/3\!-\!\alpha,x/3)-\Gamma(1\!+\!j/3\!-\!\alpha,x) \right ]  \,,
\label{Kmajor}
\end{equation}
\begin{equation}
K_{\rm destr} (x) = x^{3-\alpha} \sum_{j=0}^6
{\rm Min} (j,7\!-\!j) \,\Gamma(1\!+\!j/3\!-\!\alpha,x) \ .
\label{Kdestr}
\end{equation}
\begin{figure}[ht]
\begin{center}
\resizebox{0.7\hsize}{!}{\rotatebox{-90}{\includegraphics{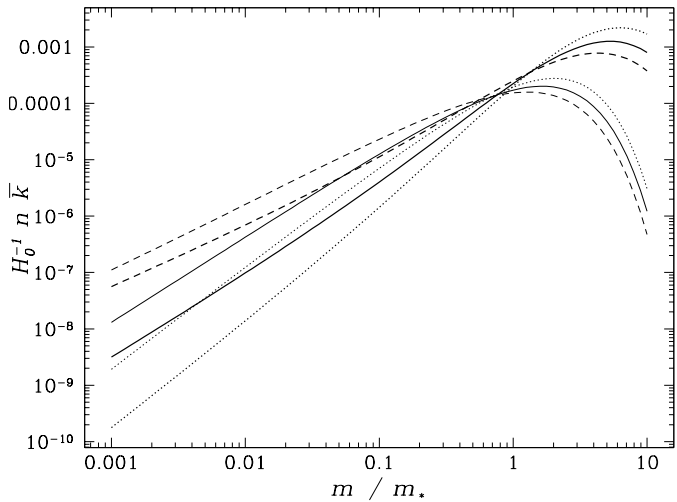}}}
\end{center}
\caption{Number of major mergers with lower mass galaxies
(eqs.~[\ref{nkofm}] and [\ref{Kmajor}], \emph{thick 
curves}) and higher mass galaxies (eqs.~[\ref{nkofm}] and [\ref{Kdestr}],
\emph{thin curves}) extrapolated to one 
Hubble time 
versus galaxy mass. 
\emph{Dotted}, \emph{solid} and \emph{dashed curves} are for $\alpha = -1.1$,
$-1.3$ and $-1.5$, respectively.
The normalization assumes $a = 12$, $v_{\rm cl} = 1000 \, \rm km \, s^{-1}$,
$n_* = 200\,n_*^{\rm field}$, with $n_*^{\rm field}
= 0.013 \,h^3 \rm Mpc^{-3}$ \citep{Marzke+98}, and $m_* = 0.1\,(M/L)_{\rm
cl}\,\ell_* = 3\times 10^{11}
h^{-1} M_\odot$.
}
\label{mergratesvsm}
\end{figure}

Figure~\ref{mergratesvsm} shows the expected number of major mergers with
smaller galaxies and 
destruction by mergers with larger galaxies that a galaxy of a given mass
should expect in a Hubble time if it sits in a typical location of a rich
cluster, assuming a constant rate in time.
The rise in merger rates at low mass reflects the rise of merger
cross-section with mass, while the decrease at high mass is caused by the
sharp decrease in the galaxy mass function yielding few galaxies to merge
with.
Figure~\ref{mergratesvsm} clearly indicates that
\emph{the probability of merger for a given galaxy is always small.}
Moreover, \emph{low and intermediate mass galaxies ($m < m_*$) are usually
cannibalized before undergoing major mergers}.


\subsection{Variation of merger rates with position in cluster}

One can go one step further and predict the variation of the merger rate with
position in the cluster.
{}From eq.~(\ref{nkofm}), the merger rate scales with radius as
\begin{equation}
{\cal R} (R,m) 
= {a \,G^2 \,m_* \,\mu^2(R) \,\rho_{\rm cl}(R) \over 16 \,
\Gamma(2\!-\!\alpha,x_m) \,v_{\rm cl}^3(R)} \,K(m/m_*) 
\ ,
\label{nkvsr}
\end{equation}
where $\mu(R) = m(r_t)/m(r_{\rm vir})$ (see eq.~[\ref{mtide}]), $m_*$ is
the mass at the break of the \emph{field} galaxy mass function and 
$x_m$ is the minimum galaxy mass in units of $m_*$.

Assuming a mass density profile $\rho \sim R^{-\beta}$ and arguing
that cluster
galaxies are severely tidally truncated by
the cluster potential as $r_{\rm gal} \sim R$ (i.e. with eq.~[\ref{finalrt}]
and assuming constant $R_p/R_a$), \cite{M92} showed that 
if galaxies also follow the
law $\rho \sim r^{-\beta}$, then their masses obey $m \sim R^{3-\beta}$.
Note that this sharp scaling of galaxy size with clustocentric distance is now
confirmed in high resolution cosmological simulations of clusters
\citep{Ghigna+98}.
Hence, \emph{the radial variation of merger rates are strongly modulated by
potential tides}.

By writing $n(R) \sim \rho(R)/m(R) \sim R^{-3}$,
I
derived
$n k \sim R^{-\beta/2}$, hence a higher merger rate inside the cluster, with a
slope agreeing perfectly with the observed elliptical fraction \citep*{WGJ93},
given $\beta = 9/4$ as predicted in early models of cluster formation
\citep{Bertschinger85}.
The derivation above has one flaw: although galaxy masses were correctly
scaled to increase with $R$, 
I
forgot to scale the fraction of
cluster mass lying within galaxies in the same way.
Therefore, one really expects $n(R) \sim \rho(R) \sim R^{-\beta}$ and 
$n k \sim R^{3-3\beta/2}$ yielding a slope $d\ln (nk) / d\ln R = -3/8$ for
$\beta = 9/4$ and a null slope for $\beta = 2$, both in disagreement with the
logarithmic gradient of elliptical fraction found by \citeauthor{WGJ93}.

One can use the more
realistic NFW density profiles 
to estimate the radial dependence of the merger rates.
An essential parameter is $\langle R_p / R\rangle$, which measures the
effectiveness of the tides from the cluster potential.
Because the dynamical friction time scales as $M/m$ times the orbital time
\citep{M95_Chalonge},  {\sl orbit circularization\/}, which to first order
operates on a
dynamical friction time scale, should be very
slow for galaxies falling onto clusters, but fairly effective for galaxies
falling into small groups.

\begin{figure}[ht]
\begin{center}
\resizebox{\hsize}{!}
{\includegraphics{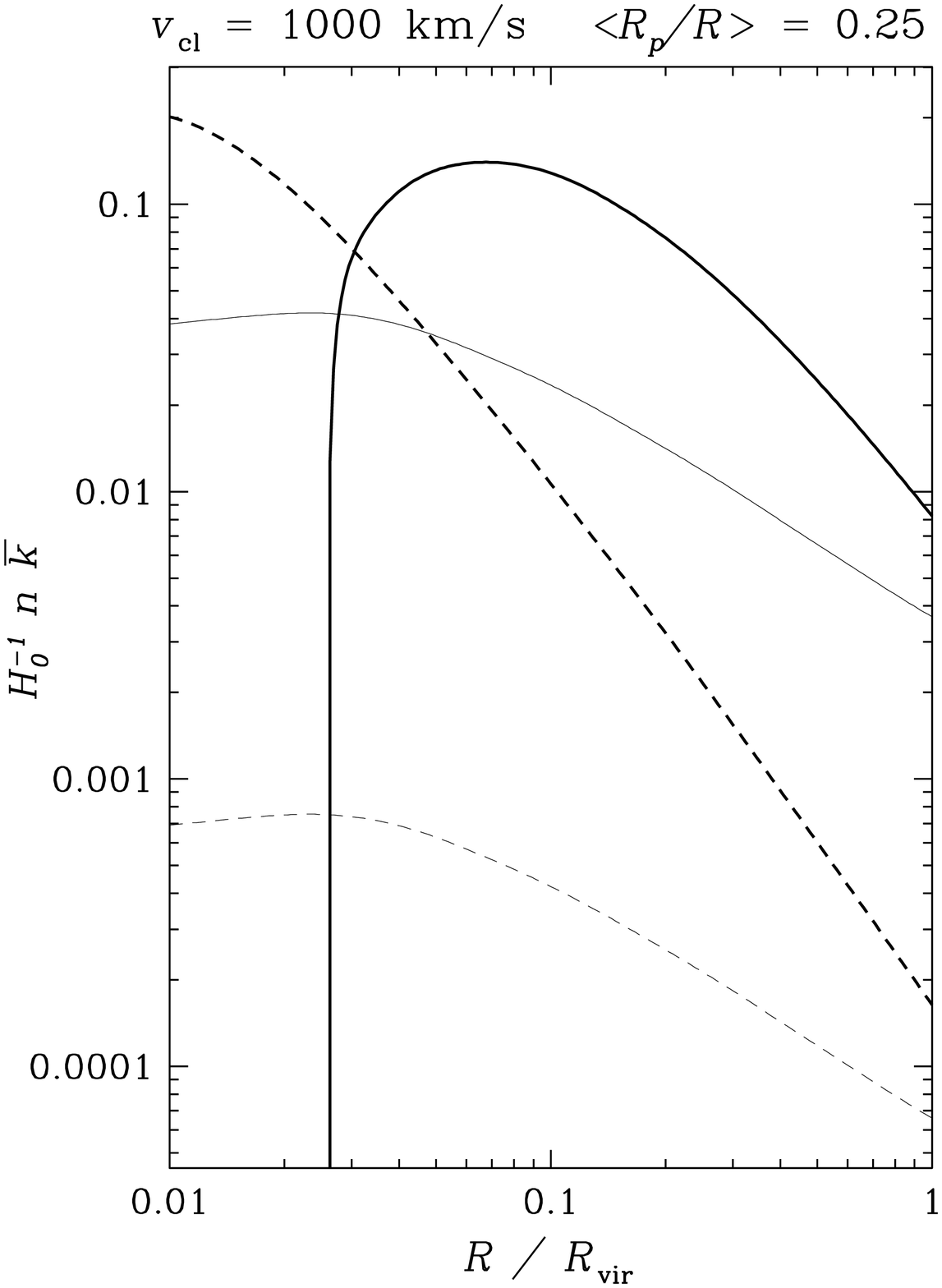}
\includegraphics{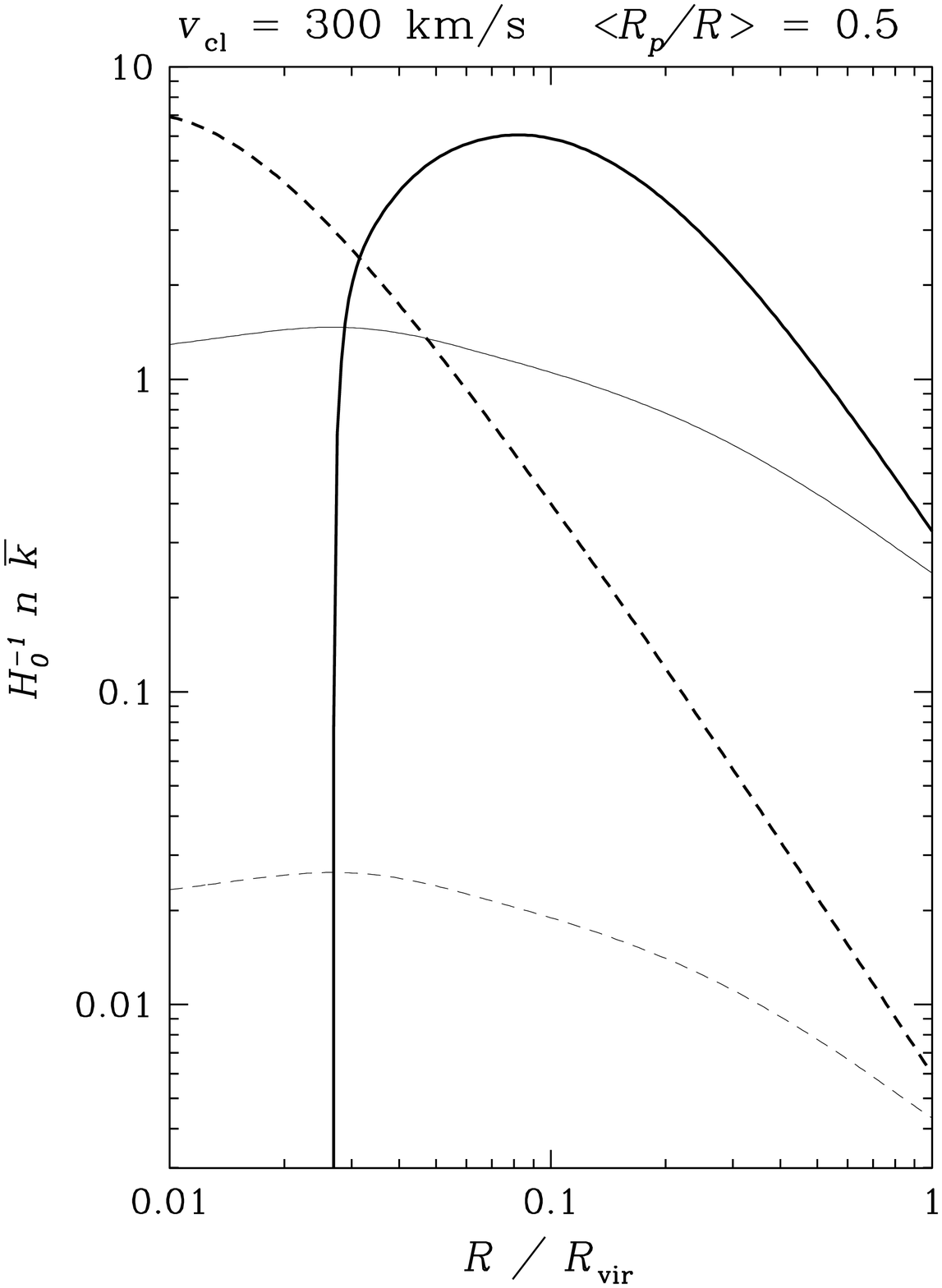}}
\end{center}
\caption{Number of major mergers a given galaxy undergoes
with lower mass galaxies (eq.~[\ref{nkvsr}], with eqs.~[\ref{nkofm}] and
[\ref{Kmajor}]), extrapolated to one 
Hubble time, versus clustocentric
radius in (\emph{left}) an NFW cluster with $v_{\rm cl} = 1000 \, \rm km \,
s^{-1}$ and (\emph{right}) a group with $v_{\rm cl} = 300 \, \rm km \,
s^{-1}$, where the galaxy
mass function has $\alpha = 1.3$.
The \emph{solid thick} and \emph{dashed thick curves} represent the expected
number of major mergers for galaxies of mass
$m = m_*^{\rm field} = \Omega_0 \rho_c / [n_*^{\rm field}
\Gamma(2-\alpha,x_m)] = 5\times10^{12} h^{-1} M_\odot$ and 
$m = 0.1\,m_*^{\rm field} = 5\times10^{11} h^{-1} M_\odot$, respectively.
The \emph{solid thin} and \emph{dashed thin curves} refer to galaxy masses
$m = m_*(R)$ and $m = 0.1\,m_*(R)$, respectively.
}
\label{mergerratevsr}
\end{figure}

Figure ~\ref{mergerratevsr} shows the predicted number of major mergers
in rich clusters and small groups, extrapolated over a Hubble time, using the
non-self-similarity of the NFW 
profiles, an exact scaling of the typical galaxy mass $m_*$ with
radius (see eq.~[\ref{mtide}]), and partial orbit circularization in groups.

The merger rates for constant mass galaxies fall off sharply at small
clustocentric radii, simply because tidal truncation of galaxies is so severe
that there are no galaxies left that are massive enough to produce a major
merger with our test galaxy.
In general, the merger rates are maximum for intermediate radii for given
test galaxy masses, and at low radii for fixed $m/m_*(R)$ (recall though 
that low mass galaxies get cannibalized before they can undergo a major
merger with a smaller galaxy).
In any event, Figure~\ref{mergerratevsr} confirms that \emph{mergers are
ineffective in clusters, but very effective in small groups}.
Note that without resorting to partial orbit circularization within groups, 
 the expected number of mergers in groups is somewhat less than
expected from 
the $v_{\rm cl}^{-3}$ scaling, because, relative to clusters, the more
concentrated NFW profiles in groups lead to stronger modulation of the merger
rate 
by the potential tides.

\section{Collisional tidal stripping in clusters and groups}

Because non-merging galaxy collisions are by essence rapid, they can be
treated as tidal shocks, and it is reasonable to assume that for tidal
features to be visible, one requires $\Delta E \geq \gamma |E|$, where $\gamma
\la 1$. Hence, $\Delta v \geq (3\gamma)^{1/2} v_g$.
Denoting $p$ and $V$ the
separation and relative velocity at pericenter, and $v_{\rm circ,p}(p)$ the
circular velocity of the
perturbing galaxy out to $p$, eq.~(\ref{dvimpulse}) leads to
a critical impact parameter 
\begin{equation}
p_{\rm crit} = {r \over (3\gamma)^{1/2}} {v_{\rm circ,p}^2 \over v_g V} \ ,
\label{ptide}
\end{equation}
where we note that $v_{\rm circ,p}$, is
almost independent of $p$ for 
realistic density profiles for the perturbing galaxy.
Then, integrating the cross-sections derived from eq.~(\ref{ptide}),
the rate of tidal interactions is
\begin{equation}
k = \langle v s(v) \rangle 
= {\pi^{1/2} \over 3 \gamma} \left ({r \over v_g} \right)^2 
{v_{\rm circ,p}^4 \over
v_{\rm cl}} \ ,
\label{ktide}
\end{equation}
and is virtually independent of the test galaxy parameters.
Integrating eq.~(\ref{ktide}) 
over perturber mass and remarking that both the galaxy and the
perturbers are tidally limited by the cluster potential, one obtains using
eq.~(\ref{elongtide}) 
\begin{equation}
n \bar k =
{\Gamma(7/3-\alpha,x_m) \over 4 \pi^{1/2} \,\gamma
\,G\,\Gamma(2\!-\!\alpha,x_m) }  \left ({v_{\rm circ} \over v_g} \right )^2 
{v_{{\rm circ},*}^4 \over m_*} \,{ \rho_{\rm cl}(R) \,\mu^{7/3} (R) 
 \over \bar \rho_{\rm cl} (R_p) \,v_{\rm cl} (R)} 
\left [ {V_p \over V_{\rm circ} (R_p)} \right ]^2 \ ,
\label{tiderate}
\end{equation}
where again $\mu(R) = m(r_t)/m(r_{\rm vir})$ (see eq.~[\ref{mtide}]),
$v_{{\rm circ},*}$ is the circular velocity at the virial radius for a field
$m_*$ galaxy.
Again, \emph{the rate of tidal encounters is independent of the galaxy mass}.
Note that the $\mu^{7/3}(R)$ dependence of the rate of tidal encounters
illustrates 
the strong modulation of this rate by the tides from the cluster potential.

\begin{figure}[ht]
\begin{center}
\resizebox{0.7\hsize}{!}{\rotatebox{-90}{\includegraphics{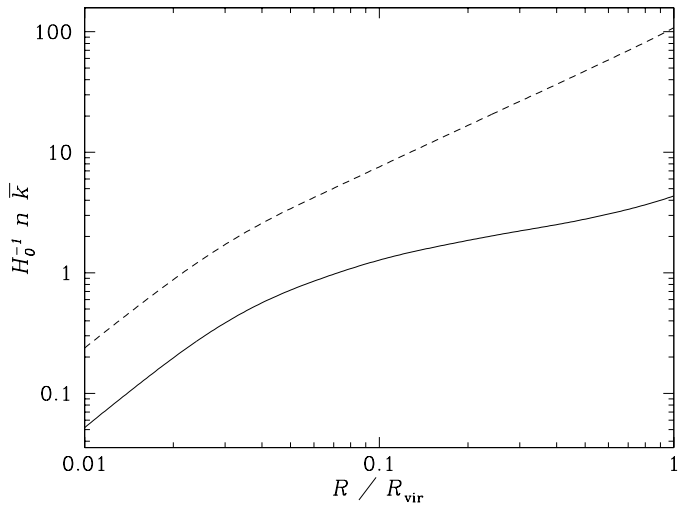}}}
\end{center}
\caption{Number of strong ($\gamma = 1/3$) tidal encounters a given galaxy
undergoes 
(eq.~[\ref{tiderate}]), 
extrapolated to one
Hubble time, versus clustocentric
radius in an NFW cluster with $v_{\rm cl} = 1000 \, \rm km \,
s^{-1}$ (\emph{solid curve}) and a group with $v_{\rm cl} = 300 \, \rm km \,
s^{-1}$ (\emph{dashed curve}). The galaxy
mass function has $\alpha = 1.3$ with field $m_* = 5 \times 10^{12} \,h^{-1}
\,M_\odot$.
}
\label{tideprof}
\end{figure}

Figure~\ref{tideprof} shows the expected (eq.~[\ref{tiderate}])
number of strong tidal collisions
for galaxies in clusters and groups, with $\langle R/R_p\rangle$ as in
Figure~\ref{mergerratevsr}. 
Although groups are preferential sites for strong tidal encounters, galaxies
in the outskirts of clusters should also witness such interactions.
However, the signature of tidal interactions lasts of order $1 \,h^{-1}\,\rm
Gyr$, so that 
the fraction of galaxies in clusters and groups 
that are currently undergoing tidal
interactions is roughly one-tenth of what is displayed in
Figure~\ref{tideprof}.

\section{Discussion}

The strong radial dependence of galaxy masses, predicted by the tidal theory
(eq.~\ref{mtide}), is clear in the cosmological simulations of
\cite{Ghigna+98}. Should we then witness \emph{inverse luminosity
segregation} in 
clusters where,
outside of the core,
galaxies become more luminous towards the
cluster periphery? 
Indeed,
\cite*{ABM98} found a weak trend of
mean galaxy magnitude versus radius for an ensemble of clusters, although they
worry that this trend is caused by observational bias. It may be that
incompleteness of the observational samples is washing out the trend rather
than creating it. 

The lack of mergers in present-day rich clusters has been noted in
cosmological simulations of clusters \citep{Ghigna+98}.
Figure~\ref{mergerratevsr} shows that in rich clusters, mergers are at best
marginally probable for high mass galaxies lying in the cluster body.
Given that high mass galaxies are rare, such merging will be difficult (but
not impossible) to
detect observationally or in simulations.




{}From their H$\alpha$ prism surveys of galaxies in clusters,
Moss and co-workers \citep*{MW93,MWP98,BM98} note $\simeq 30\%$ of spiral
galaxies in rich clusters exhibit a \emph{compact} H$\alpha$ morphology and
roughly half of these tend to be
morphologically disturbed \emph{and} have nearby neighbors.
Another half of these compact H$\alpha$ emission galaxies are in the cluster
core, and presumably those have no tidal companions, suggesting that they are
harassed by the cluster potential.
But the first half, outside the core are probably \emph{bona fide} cases of
strong tidal interactions within clusters, leading to an overall frequency of
20\% of all galaxies outside the cores of clusters.
It remains to be seen if they are
associated to substructures such as infalling groups.
Correcting Figure~\ref{tideprof} for the 1 Gyr duration of these tidal features
produces an absolute frequency of tidally interacting galaxies in clusters
in rough agreement with that found by Moss and co-workers.

\end{document}